# The importance and impact of discoveries about neural crest fates

Heather C. Etchevers[1,*], Elisabeth Dupin[2] and Nicole M. Le Douarin[2,*]

[1] Aix Marseille Univ, INSERM, MMG, Faculté de Médecine AMU, 27 boulevard Jean Moulin 13005 Marseille, France. (Phone : +33 4 91 32 49 37)

[2] Sorbonne Universités, UPMC Univ Paris 06, INSERM, Institut de la Vision, UMR S 968, 17 rue Moreau, 75012 Paris, France. (Phone : +33(0)1 53 46 25 37)

* Corresponding authors:
nicole.ledouarin@academie-sciences.fr
heather.etchevers@inserm.fr

Abbreviations: Abbreviations:

DRG, dorsal root ganglion; E, embryonic day; EDN, endothelin; ENS, enteric nervous system; HSCR, Hirschsprung disease; MEN, multiple endocrine neoplasia; NC, neural crest; PNS, peripheral nervous system; QCPN, Quail non Chick Peri-Nuclear; r, rhombomere; SCP, Schwann cell precursor; WS, Waardenburg syndrome.

Ms content: Main text, 6032 w/o ref; Table1, Figures 1 and 2; Timeline

# Summary

**We review here some of the historical highlights in exploratory studies of the vertebrate embryonic structure known as the neural crest. The study of the molecular properties of the cells that it produces, their migratory capacities and plasticity, and the still-growing list of tissues that depend on their presence for form and function, continue to enrich our understanding of congenital malformations, pediatric cancers but also of evolutionary biology.**

# Introduction

Wilhelm His is credited with the first description of the neural crest, from careful observations of chick embryos under the microscopes of the era. He had noticed, on the dorsal side of the closing neural tube, the formation of a *cord* of cells lying *between* the superficial ectoderm and the underlying neural tube, which he called *Zwischenstrang* (His, 1868). Furthermore, His described cells emerging from it, gliding over the surface of the neural tube, aggregating laterally and forming the spinal ganglia. The initial area where the lateral edges of the neural plate folded and adjoined was therefore first called the "ganglionic crest" and later came to be known as the "neural crest". His' discovery, although not readily accepted at that time by his colleagues, was of great significance, as will be shown in this overview of what turned out to be a major component of the vertebrate embryo.

The reason for which the NC had been mostly ignored by most embryologists for over half a century was its cardinal characteristic, which is to be transitory. The NC population forms, depending on the species, approximately when the lateral borders of the neural plate (also called "neural folds") join dorsally to close the neural tube. Thereafter, NC cells undergo an epithelial-to-mesenchymal transition, migrate away from their source to invade the entire developing embryo, and settle in elected sites, where they develop into a large variety of cell types.

As they disperse, NC cells are usually indistinguishable from those of the embryonic tissues through which they move. Our knowledge about the fate and contributions of NC cells to the tissues and organs of the vertebrate body was acquired through various experimental strategies. The first consisted in selective ablation of segments of the neural tube or neural folds preceding NC cell emigration, followed by observing which subsequent structures were missing. Techniques were devised later to track the NC cells during their migration to their final destinations and ultimate differentiation. Such techniques, broadly used in developmental biology, are collectively known as "lineage tracing" or "fate-mapping". The many roles of NC cells in vertebrate physiology could then be demonstrated; they become an astonishing variety of cell and tissue types all over the body in all vertebrates, including humans.

The implications of NC cell diversity and the knowledge of signaling pathways and regulatory genes in NC development have been key to understanding the origins of many human disorders. These sometimes have seemingly unconnected phenotypes spread throughout the body.

The aim of this article is to provide a brief overview of NC research from its discovery to the present, together with some significant examples of how understanding NC cell biology has enriched other fields, such as evolutionary biology, behavioral biology and medicine.

## Recognition of the neural crest as a distinct embryonic structure

Establishing the existence of the NC as a distinct embryonic structure was for many decades the subject of active controversy, for which its transitory nature and the fact that its constitutive cells undergo extensive migration throughout the developing body were responsible. It took over eighty years to settle the debate, when Sven Hörstadius stated in his landmark monograph that "[i]t seems clearly established today [in 1950] that the neural crest forms a special rudiment already present in the open neural plate stage" (Hörstadius, 1950).

As mentioned above, His' views were far for being adopted by the embryologists of that time. As examples, Balfour (1881; Balfour and Foster, 1876) and Kastchenko (1888), among others, thought that spinal nerves were outgrowths of the spinal cord and not at all produced by the NC; moreover, the spinal ganglia were supposed to originate from the mesoderm, produced by the myotomes.

Curiously, the capacity of this ectodermal structure to yield mesenchymal cells was one of its first derivatives to be readily identified in fish. The same Kastschenko, working on Selacian development, concluded that some of the head mesenchyme originated from the cephalic NC while Goronowitsch (1892; 1893) recognized, from his studies on teleost embryos, that the cephalic NC was providing the head with skeletogenic mesenchyme, though he also denied a NC origin for the spinal ganglia.

The intense debates generated by these discrepancies could have been resolved by the work of Julia Platt (1893; 1897) - one of the rare women involved in science at that time- who found that not only cranial ganglia and nerves but also mesenchymal cells participating in bones and cartilages of the visceral arches, as well as the dentine of the teeth, were of NC origin. She even created the term of *mesectoderm* for the mesenchyme of ectodermal origin that she distinguished from the *mesentoderm* derived from the mesodermal germ layer.

Investigations were further pursued on other vertebrate species and the conclusions of the different authors still differed: some supported the ectodermal origin of mesenchyme from the NC while others did not accept that mesenchymal cells could be produced by the ectoderm (eg, Holmdahl, 1928; Landacre, 1921; Stone, 1922). The roots of this controversy grew out of von Baer's "germ layer" theory (1828), according to which homologous structures within the same animal and across different animals must be derived from material belonging to the same germ layer (ectoderm, mesoderm or endoderm). According to this view, mesenchymal (loose) cells, and specifically the skeletal tissues differentiating therefrom, could arise only from mesoderm. It was only until the nineteen twenties that thorough observations, associated to experimental work carried out essentially in amphibian embryos (Landacre, 1921; Stone, 1922; Stone, 1926), led to the demonstration that a proportion of mesenchymal cells of the vertebrate embryo are ectodermally derived via the NC.

During the first half of the twentieth century, investigations concerning NC cells were conducted essentially with amphibian and fish models, leading to demonstrations of the contribution of the NC to its other derivatives, such as the peripheral nervous system (PNS). This was achieved thanks to the development of lineage tracing strategies to follow the fate of NC cells after they had left the neural primordium and mingled with cells and tissues of other origins.

## Following neural crest cell migration

Destruction or microsurgical removal of the neural tube or neural folds *in situ* was widely used as an approach to identify potential NC contributions (Dushane, 1938). It was helpful, for example, in interpreting the common origin of thymic and cardiac outflow tract malformations in human DiGeorge syndrome, since they can be phenocopied in large part by removing NC at the level of the posterior hindbrain (Bockman and Kirby, 1984), and in identifying that forebrain and pituitary growth depends on NC cells from the level of the midbrain (Creuzet, 2009; Etchevers et al., 1999). However, extirpation of embryonic territories at early stages can sometimes trigger regenerative responses, which are able to compensate for certain types of ablations. Such techniques can establish that a primordium is necessary for the formation of a given structure, but not that it is sufficient or even contributes directly to its tissues.

Labeling distinct regions of the neural primordium *in situ* with vital dyes (that colored but did not poison cells, such as Nile Blue or Nile Red) enabled and confirmed important discoveries initiated by ablation approaches. For example, in addition to the spinal ganglia, NC cells were shown to be involved in the production of the pigment cells, of the cells that line nerves throughout the entire PNS, and to contribute to the teeth and to the dorsal fin (in fish) (Dushane, 1938; Twitty and Bodenstein, 1941). This technique was not entirely reliable, due to dye diffusion over time, but there were good examples of its efficacy, such as the thorough studies performed by Hörstadius and Sellman (1941; 1946) on the contribution of NC cells to amphibian pharyngeal arch cartilage.

Over forty years later, labeling cells with fluorescent vital dyes brought additional contributions through a painstaking method pioneered by Marianne Bronner and Scott Fraser, which consisted in the microinjection of a large fluorescent carbohydrate into single cells to visualize NC cells and their progeny without issues of membrane-to-membrane diffusion (Bronner-Fraser and Fraser, 1988, 1989; Collazo et al., 1993; Fraser and Bronner-Fraser, 1991). In combination with novel techniques of imaging, this method opened many windows of opportunity to follow the dynamics and fate decisions of NC cells *in vivo*. In addition to the chick, it was used in the zebrafish embryo (Raible and Eisen, 1994; Schilling and Kimmel, 1994) as well as early studies of mammalian NC fate using mouse embryo culture (Serbedzija et al., 1994). By following single cells until such time as the dye was diluted by cell divisions, understanding accrued about their early potential and migratory behavior, thus demonstrating for the first time, *in ovo* multipotency of individual premigratory NC cells (Bronner-Fraser and Fraser, 1988). However, this approach could not be generalized to the behavior of their neighbors or to the integration of labeled cells into late functional structures.

Injecting *LacZ*-expressing, infectious but replication-incompetent retroviruses was a complementary approach to study multiple separate cells, depending on the concentration of the initial solution (Mikawa et al., 1991). The virus solution could be placed within the neural tube, to concentrate efforts on identifying individual NC derivatives within a specific distant destination, such as the heart (Boot et al., 2003) or dorsal root ganglia (DRG) (Frank and Sanes, 1991), or within the somites (Epstein et al., 1994). This labor-intensive approach led to the *in vivo* identification of bipotent neuron/glial NC precursors in the DRG (Frank and Sanes, 1991); however it presented the technical drawback of not being able to control the extent of the initial infection and was not widely adopted in studies of NC cells.

## Construction of embryonic chimeras: exploiting graft-specific cell properties

Raven (1931) was the first to trace the fate of NC cells by exchanging defined fragments of the neural folds between *Triturus* (salamander) and *Ambystoma* (axolotl). The identification of host versus donor cells, based solely on differences in the size of their nucleus, was highly labor-intensive. Nonetheless, this method permitted the origin of populations of cells in chimeric tissues to be recognized (single-cell resolution was not possible). Thanks to this approach, the NC origin of dental papillae – the mesenchyme below each tooth bud that gives rise to dentine and pulp – was discovered in amphibians nearly 70 years before its confirmation in mammals (Raven, 1935).

Significant progress in terms of lineage tracing was made through the use of autoradiographic techniques following injection of tritiated ($^3$H)-thymidine to label dividing neuroepithelial cells (Sauer and Walker, 1959, in chick; Sidman et al., 1959, in mouse). This technique allowed grafts of defined neural territories labeled in a donor embryo to be implanted in an unlabeled recipient of the same species. It was initiated by Jim Weston to follow the migration of chicken NC cells (Weston, 1963) and later used by Johnston (1966) and Noden (1975), and by Chibon (1964; 1967) in *Pleurodeles* (a salamander). $^3$H-thymidine labeling, however, suffers a few major limitations: it is unstable, being diluted along cell cycles if the grafted cells proliferate; concentration of the radioactivity in nuclear DNA can cause toxic effects or alteration of cell behavior; moreover, in case of grafted cell death, dividing host cells could potentially take up the released radioactive DNA. Nonetheless, several important results can be credited to this technique, such as the confirmation of the NC origin of spinal and sympathetic ganglia, and pigment cells (Weston, 1963). The derivation of Schwann cells from a mesodermal source was definitively excluded, and most of them were recognized as NC-derived at this time. Using $^3$H-thymidine-labeled *Pleurodeles* NC cells implanted into an unlabeled host, Chibon (1967) also confirmed the NC origin of major components of the head skeleton, as did Johnston (1966) in the chick, although the possibility of a ventral neural tube origin for Schwann cells along motor spinal nerves was considered.

## The quail-chick chimera technique led to a comprehensive avian NC fate map

The quail-chick marking system provided a means to study the development of the NC in amniote vertebrates, which had not been as often used in experimental embryology as amphibians or fish. It was designed by one of us (Le Douarin, 1969) following the observation that cells of the quail (*Coturnix coturnix japonica*) could be readily distinguished from those of the chick by the structure of their nucleus. Quail nuclei are characterized by the presence of a large nucleolus at all life stages, which is caused by a mass of heterochromatin associated with the nucleolus (essentially made up of RNA). This particularity is rare in the animal kingdom (Le Douarin, 1971b) and does not exist in the chick (Le Douarin, 1969; Le Douarin, 1971a). When quail cells are transplanted into a chick embryo or associated with chick tissues *in vitro*, the cells of each species retain their nuclear characteristics and can be readily distinguished at the single-cell level in chimeric tissues after using the Feulgen-Rossenbeck histological nuclear stain, provided that the section includes the nucleolus (Le Douarin, 1973). This method of detection has been widely employed to analyze cell fates over time. A significant improvement in the identification of the cells of the two species in the chimeras occurred with the production of monoclonal antibodies able to recognize species-specific antigens expressed by one but not the other species. The antibody against a peri-nuclear antigen, QCPN (for Quail non Chick Peri-Nuclear), prepared by B. M. Carlson and J. A. Carlson from the University of Michigan, turned out to be particularly useful to analyze quail-chick chimeric tissues from mid-1990s onward.

The fate of NC cells in the avian embryo was systematically investigated by constructing chimeras in which a fragment of the neural tube (or neural fold at the head level) of a chick embryo was removed prior the onset of NC cell emigration at this level of the neural axis and replaced by its exact counterpart from a quail embryo at the same developmental stage. The same types of grafts were initially performed in both directions, i.e. from quail to chick and from chick to quail, to validate the approach. The stability of cell labeling allowed migration and fate of NC cells to be followed during the entire incubation period and even for a time after birth (e.g. Kinutani and Le Douarin, 1985).

This exploration of the fate and migration pathways of NC cells was effective in determining which cell types came from the NC in birds and from which axial levels of the neural primordium (for references, Le Douarin, 1982; Le Douarin and Kalcheim, 1999) (**Table 1**). Together, these experiments provided a fate map of NC cells regarding production of mesenchymal cells, pigment cells and of the sensory, autonomic and enteric components of the PNS (**Figure 1**). New endocrine derivatives, such as the carotid body and the calcitonin-producing cells of the ultimobranchial body, were thus identified (Le Douarin and Le Lièvre, 1970; Le Douarin et al., 1972; Le Douarin et al., 1974; Pearse et al., 1973; Polak et al., 1974), and the contribution of the NC to the cephalic structures was, for the first time, fully demonstrated. (e.g. Le Lièvre and Le Douarin, 1975; reviewed in Le Douarin and Kalcheim, 1999).

The precise levels of the neural axis from which the various derivatives of the NC originate were determined. It turned out that defined regions of the spinal cord and medulla oblongata (posterior hindbrain) were dedicated to producing the sympathetic, parasympathetic or enteric ganglia (**Figure 1**). For example, the so-called "vagal" level of the NC, extending between the axial levels of somites 1 to 7 and corresponding to the level of emergence of the vagus nerve, provides the entire gut with NC cells that, later on, differentiate into the two plexuses of the enteric nervous system (ENS; Le Douarin and Teillet, 1973). Some of these cells (from pre-somitic levels to somite 3) also contribute to the perivascular walls of the posterior pharyngeal arteries and septation of the outflow tract of the heart (Arima et al., 2012; Kirby et al., 1983; Le Lièvre and Le Douarin, 1975) (**Figure 2**). *Such work also revealed a so far unknown "principle" that both pre- and post-ganglionic neurons of the ENS originate from the same level of the neural axis.*

This identification of common lineages between very disparate tissues shed light on the pathophysiology of clinical associations that had been unexplained to that point, a group of diseases for which the term "neurocristopathy" was coined (Bolande, 1974; Bolande, 1997; reviewed by Etchevers et al., 2006). In the above example of the "vagal" NC, developmental biology provided the basis for understanding that lack of distal development of the ENS at the level of the rectum and colon, giving rise to Hirschsprung disease (HSCR), could stem from a genetic problem to which migratory vagal NC would be particularly susceptible. In 1994, HSCR phenotypes due to mutations in genes whose function is necessary for NC migration into the developing gut were first identified in human patients and in several mouse genetic models (reviewed by Bondurand and Southard-Smith, 2016, Table 1). For example, mutations in *RET*, a gene encoding a receptor for glial-derived growth factor can cause this congenital intestinal aganglionosis (Edery et al., 1994; Iwashita et al., 2003). Such is also the case for loss-of-function mutations in the genes encoding the cytokine endothelin-3 (*EDN3*) and its type-B receptor (*EDNRB*) (Edery et al., 1996; Hofstra et al., 1996; Hosoda et al., 1994; Puffenberger et al., 1994).

HSCR can be sometimes associated with pigment cell (melanocyte) defects, dysmorphic facial features, auditory impairment and occasionally, cardiac outflow tract anomalies in the clinically heterogeneous Waardenburg syndrome (WS). In piebaldism, a congenital absence of melanocytes in areas of the skin, often affecting the midline at the scalp-forehead interface and leading to a white forelock, can be associated with auditory problems. These problems arise following the lack of melanocyte colonization of the inner ear, where they are needed for cochlear function (Steel and Barkway, 1989). The entire constellation of WS symptoms, including piebaldism, can result from the mutation of a handful of individual genes, such as the transcription factors *SOX10* (Pingault et al., 1998), *PAX3* (Dow et al., 1994) and *SNAI2* (Sanchez-Martin et al., 2002). These factors have been characterized in the chick embryo as key players in the gene regulatory networks that control the formation and migration of NC cells (Sauka-Spengler and Bronner-Fraser, 2008; Simoes-Costa and Bronner, 2015). Depending on their clinical severity and association with other symptoms into syndromes, different classes of WS have been distinguished and the genetic origins for many but not all of these classes have been discovered. To add to the complexity, they are not always simple monogenic conditions but those genes identified are always important to NC development. The neurocristopathy concept helps explain such phenotypic diversity (Etchevers et al., 2006; Vega-Lopez et al., 2018).

Both ablation experiments and quail-chick chimeras demonstrated an additional contribution to the ENS from the lumbosacral NC, located caudal to the level of somite 28 (Burns and Douarin, 1998; Le Douarin and Teillet, 1973; Yntema and Hammond, 1955). Thus, in birds, the vagal and lumbosacral extremities of the spinal cord contain the ganglionic neurons of the entire ENS (**Figure 1**). A third region of the NC, located between somites 18 and 24 (at the level of and just posterior to the forelimb), provides the adrenal gland with adrenaline- and noradrenaline-producing cells, together with the sympathetic ganglia corresponding to this axial level, with no contribution to the ENS. More posterior levels of the NC generate the accessory aggregates of catecholaminergic cells that develop along the dorsal aorta (Le Douarin and Teillet, 1974). This sympathoadrenal lineage is particularly susceptible to developing into pediatric cancers such as neuroblastoma and pheochromocytoma, a cancer of the adrenal medulla. It can also be the site of a heritable genetic predisposition to multiple endocrine neoplasias of type II (MEN2) including pheochromocytoma as well as tumors of the thyroid medulla (involving the calcitonin-producing cells), carotid bodies and other paragangliomas. Like malformation-type neurocristopathies, genetic and phenotypic heterogeneity is the rule rather than the exception. Interestingly, mutations in some of the same genes can cause either malformations or cancers in NC-derived tissues. Such is the case for the growth factor receptor *RET*, which can cause HSCR with loss-of-function mutations but MEN2 with gain-of-function mutations (Takahashi et al., 1999). Local ("somatic") mutations within NC-derived cells of genes known from receptor-activated signaling pathways are also mobilized in multiple cancer types. Many of these genes, known as oncogenes, not only are associated with cancer but also cause birth defects in the neurocristopathy spectrum, such as the giant congenital melanocytic nevus or cardio-facial-cutaneous syndrome (e.g. Etchevers et al., 2018; Niihori et al., 2006).

## What about the study of NC in mammals?

Embryonic manipulation being technically challenging in mammals, solutions to trace lineages in mice are based on the endogenous or chemically inducible expression of genes in the desired cell population. Such experiments must be carefully designed, since they often rely on the specificity of regulatory elements to drive gene expression in a given tissue and a given moment of development; such elements can be re-activated during co-option of the gene's function at other moments in life, confounding interpretation, or the expression of even one allele with the marker can interfere with a necessary gene function. The development of conditional site-specific recombination, by combining an allele driving the expression of a bacteriophage-derived *Cre* recombinase with an allele containing a *loxP*-flanked (floxed) sequence to be excised in the target cell (Sauer and Henderson, 1988), was as beneficial to the study of the NC as it was to developmental biology as a whole. A particularly influential transgene construction was devised by McMahon and colleagues containing enhancers that drive *Wnt1* gene expression in the dorsal neural tube and, thereby, in premigratory NC in the mouse (Danielian et al., 1998). A mouse line, commonly known in the literature as *Wnt1-Cre*, was then created.

Interpretation of conditional lineage tracing in mice relies on the spatiotemporal specificity of *Cre* driver expression. The use of *Wnt1-Cre* to target most or all the premigratory NC cells – perhaps not putative late-emigrating NC cells nor the small NC subpopulation from the diencephalic neural folds rostral to the anterior limit of endogenous *Wnt1* expression – has still contributed a great deal to the understanding of mammalian NC derivatives. *Wnt1-Cre*- labeled NC cells yield the same head musculoskeletal elements as in birds as well as ocular and periocular tissues, and, in addition, teeth and palate (Chai et al., 2000; Gage et al., 2005; Jiang et al., 2002; Matsuoka et al., 2005). They are also found both around and within the cardiac outflow tract and tricuspid valves (e.g. El Robrini et al., 2016; Jiang et al., 2000) as well as in the skin, at the level of innervated hair follicles and glands (Wong et al., 2006). The permanent nature of marker expression after recombination allowed for new, perhaps late sites of NC colonization to be identified, such as the marrow of adult long bones (Nagoshi et al., 2008).

However, the *Wnt1-Cre* allele activates ectopic Wnt signaling in the ventral midbrain, perhaps because of the confounding genomic rearrangements induced by insertion of the transgene (Goodwin et al., 2017; Lewis et al., 2013). Other *Cre* lines have been developed to also drive target allele recombination in some or most NC cells to examine distinct aspects of their development, *Sox10-Cre*, *htPA-Cre* and *P0-Cre* among others, which label migratory, not premigratory NC cells (Matsuoka et al., 2005; Pietri et al., 2003; Yamauchi et al., 1999). Further refinement of the system has included the development of "inducible" conditional alleles that respond to the exogenous administration of a drug, to bypass those phenotypes restricted to the first time a gene influences a biological process and examine recombination in the lineage at desired moments. Additional advantages of the Cre-lox system, largely exploited over the last fifteen years, have been to combine such analyses of lineage tracing with the responses of the same cells to genetic gain- or loss-of-function, but we will leave the exploration of these many recent studies to our readers' diligence.

## Neural crest stem cells

The wide diversity of cell types arising from NC cells and their extensive migration around the body are features in common with hematopoietic stem cells. These give rise to all blood cell lineages throughout life and were the first tissue-specific stem cells identified in higher vertebrates (Bradley and Metcalf, 1966; Till and McCulloch, 1961; reviewed by Metcalf, 2007). Inspired by these findings, NC developmental potential began to be studied at a cellular level in the late 1980s (see **Timeline**). Cephalic quail NC cells were first explanted just after having left the mesencephalic neural folds, then replated individually under conditions suitable for the differentiation of many principal NC-derived phenotypes (Baroffio et al., 1988). Clones derived from single cells comprised various combinations and numbers of glia, adrenergic neurons and melanocytes, together with less frequent cartilage cells (Baroffio et al., 1988; Baroffio et al., 1991; Dupin et al., 1990). Using a similar *in vitro* clonal assay, stem cells producing both neurons and glia in the mammalian embryonic CNS were also discovered around this time (Temple, 1989). Further investigations in quail showed that cephalic NC cells are highly multipotent and can produce peripheral neurons and glia, melanocytes, chondrocytes and osteocytes *in vitro* (Calloni et al., 2007; Calloni et al., 2009). These data support a common cellular origin for components of the cranial PNS and craniofacial mesenchymal tissues in vertebrates, leading to a stem cell lineage model of NC cell diversification (Dupin et al., 2010; Trentin et al., 2004).

In the trunk NC, pioneer studies had demonstrated bipotency in avian NC cells that had been isolated after migrating from neural tubes explanted just before NC cell delamination (Cohen and Konigsberg, 1975; Sieber-Blum and Cohen, 1980). Variations on this simple and efficient technique of NC cell isolation were later employed for many *in vitro* studies of avian and mammalian NC cells, including human (Thomas et al., 2008; reviewed in Etchevers, 2011). Together with improvements in culture conditions, availability of antibodies against a wider range of phenotype-specific antigens in more species, a more comprehensive picture of the developmental repertoire of NC cells has emerged (reviewed in Dupin and Sommer, 2012; Dupin et al., 2018). Some avian trunk NC cells are multipotent progenitors for glial cells, autonomic neurons, melanocytes and smooth muscle cells, while others are only bipotent (Lahav et al., 1998; Trentin et al., 2004). In the rat, NC stem cells can clonally generate glia, autonomic neurons and smooth muscle cells and are able to self-renew in culture (Stemple and Anderson, 1992). Similar progenitors have also been found in the fetal rat sciatic nerve (Morrison et al., 1999). The cardinal stem cell property of self-renewal is widely observed in individual, oligopotent NC cells *in vitro* across species: in bipotent quail NC progenitors for melanocytes/glial cells (Trentin et al., 2004) or glial/smooth muscle cells (Bittencourt et al., 2013); in autonomic/sensory neuron progenitors of the early murine NC (Kléber et al., 2005), and in premigratory chick cranial NC cells, maintained as multipotent floating spheres *in vitro* (Kerosuo et al., 2015).

It is remarkable that multipotency of NC-derived progenitors can extend into adulthood. At postnatal and adult stages, genetic lineage tracing has shown in rodents that not only the PNS, but also hair follicles, cornea and iris, carotid body, dental pulp, bone marrow and oral mucosa contain oligo- or multipotent NC cells with self-renewal ability in culture (reviewed in Dupin and Coelho-Aguiar, 2013; Motohashi and Kunisada, 2015). Such widespread tissue distribution of multipotent NC cells may be maintained through their association with the pervasive network of PNS nerves and nerve endings throughout the body. NC cells assume a Schwann cell precursor (SCP) identity while located in their "niche" along the nerve; however, they can detach from it and adopt non-glial fates both during development and lifelong regenerative turnover such as found in hair follicles and the rodent incisor (reviewed in Furlan and Adameyko, 2018). This was first shown *in vivo* for melanocytes (Adameyko et al., 2009), but such SCP can also give rise to cranial parasympathetic neurons (Dyachuk et al., 2014; Espinosa-Medina et al., 2014), endocrine chromaffin cells (Furlan et al., 2017), osteoblasts and dental pulp cells (Kaukua et al., 2014), and enteric neurons (Espinosa-Medina et al., 2017). SCP are likely to be the resident NC-derived mesenchymal stem cells of the adult mouse bone marrow and body dermis (Fernandes et al., 2004; Isern et al., 2014; Nagoshi et al., 2008).

Since ethical and scientific challenges make it difficult to isolate and study embryonic human NC cells (Thomas et al., 2008), several protocols to derive NC-like cells from pluripotent human embryonic stem cell lines (Kerosuo et al., 2015; Lee et al., 2007), induced pluripotent stem (iPS) cells (Avery and Dalton, 2015; Menendez et al., 2012) or even by direct reprogramming of pediatric fibroblasts (Kim et al., 2014) have been developed. Combined with genome-wide analysis of gene expression, these cellular models can help define human NC cell differentiation mechanisms during normal development and pathology, such as sequential enhancer activation and epigenetic annotations (Bajpai et al., 2010; Rada-Iglesias et al., 2012). Patient-derived NC-like cells may contribute to better understanding of human neurocristopathies, provide a cellular platform for drug screening, and permit innovative therapies based on autologous cell grafts and tissue engineering (Liu and Cheung, 2016).

## Contributions of the neural crest to the vertebrate head and their significance

The above described experimental approaches contributed to understanding the wide range of NC-derived tissues and organs in the head. The role of the mesectoderm was found to be quantitatively far more important for head development than previously thought before the 1970s (**Figure 2**). Tissues of NC origin in the head are not only skeletogenic for the entire facial skull, inner ear and rostral brain case, but also include soft tissues such as the dermis, fascia and tendons of the facial and ocular muscles, interstitial cells in all head and neck glands examined, and adipocytes of the face and neck (Billon et al., 2007; Couly et al., 1993; Grenier et al., 2009; Le Lièvre and Le Douarin, 1975; Noden, 1978) (**Table 1**). Strikingly, the striated myocytes of avian iris muscles (Nakano and Nakamura, 1985), but also the ciliary bodies and the corneal endothelium and stroma, are all specialized derivatives of the cephalic NC in both chicken (Creuzet et al., 2005; Johnston et al., 1979) and mouse (Gage et al., 2005).

Cephalic NC cells are also the source of vascular smooth muscle in a restricted area of the circulatory system (**Figure 2**). This "branchial" sector extends distally from the cardiac outflow tract, whose septation depends on the presence of NC cells (Kirby et al., 1983; Waldo et al., 1998), along the great arteries and jugular veins, where they lie in the ventral neck and face, out to the capillary beds and choroid plexuses of the face and forebrain (Bergwerff et al., 1998; Couly and Le Douarin, 1987; Etchevers et al., 1999; Etchevers et al., 2001). NC cells also give rise to both the fibrous dura mater and the highly vascular pia mater of the forebrain (only) meninges, although not to their vascular endothelium (Couly and Le Douarin, 1987; Le Lièvre and Le Douarin, 1975); the pericytes accompanying penetrating capillaries into this part of the brain and the eyes are NC derivatives (Etchevers et al., 2001), while in other parts of the brain and spinal cord, meninges and pericytes appear to be of mesodermal origin (Couly et al., 1992; Couly et al., 1995). Such regionalization and invisible differences in functional potential within the cardiovascular system are relevant to the reproducible ocular, brain and facial territories affected by Sturge-Weber and cerebrofacial arteriovenous metameric syndromes, which can also comprise nerve and bone malformations (Krings et al., 2007; Nakashima et al., 2014).

The neck and shoulder area has been shown in the mouse to be the interface between two different territories. In the head and much of the neck, NC ensures muscle tendon attachment points to specific skull and neck bones. In the rest of the body, tendons come from the mesoderm. The NC origin of the connective tissues of tongue and pharynx constrictor muscles, for example, explained certain symptoms associated with malformation syndromes like Klippel-Feil or cleidocranial dysplasia (Matsuoka et al., 2005). Furthermore, the classification of muscles according to their NC-derived or solely mesodermal connective tissue, led to an improved strategy for asserting bone homologies in the fossil record.

## Concluding remarks

The embryological results uncovering the paramount role of the NC cells in building the various head tissues, together with other considerations, led Gans and Northcutt (1983) to develop the concept of the vertebrate "New Head", according to which properties of placodal sensory organs and the newly acquired NC enabled the evolutionary transition from protochordates to vertebrates. The development of a head as we know it – with protection by the skull, prominent jaws and highly developed sensory organs and brain – coincided with changes in animal lifestyle, from filter-feeding animals to vertebrates endowed with active predation and locomotion and sophisticated sensory-motor modalities. This theory was updated to account for important experimental embryological findings in the interim (Northcutt, 2005) and has been largely accepted.

The plasticity of the NC lineage may contribute not only to natural selection but also to the effects of human selection, through domestication. During selection for docility, many species of domesticated animals have also acquired converging features, including a white forelock, floppy ears, large forehead, and smaller jaws and teeth (reviewed by Trut et al., 2009 and Wilkins et al., 2014). In so doing, breeders have also sometimes unwittingly selected for the attendant associated pathologies. White cats with blue eyes are deaf, due to absence of melanocytes from their cochlea (Mair, 1973). The *Splotch* mouse is a spontaneous mutant in a transcription factor gene essential for NC migration and melanocyte differentiation, among other processes; heterozygotes develop white coat splotches on the belly. These mice are subject to cardiac outflow tract defects and HSCR, like their human counterparts with WS (Conway et al., 1997; Serbedzija and McMahon, 1997). Selection for tameness correlates with endocrine effects as well; NC cells of the adrenal medulla secrete epinephrine (adrenalin), while domesticated animals exhibit less reactivity and stress in the presence of humans than their wild counterparts.

In summary, there may well be another 150 years of discoveries to be made about the intriguing and adaptable NC cell population, to the continued benefit of both medicine and science.

**Table 1: Derivatives of the neural crest**

**Table 2: Timeline (text)**

**Figure 1: Fate map of NC derivatives**

Representations of chicken embryos at 7 and 28 somite pairs (S), with neural folds capable of generating neural crest (NC) cells in gray, indicating in colors the axial levels from which different classes of NC derivatives originate. Adapted from (Le Douarin et al., 2004).

**Figure 2: Migration of the cephalic NC cells to the craniofacial and cardiac regions**

A. Schematic dorsal view of chicken embryo with 5 somite pairs, at about 30h incubation. Colors represent original levels of NC cells in the neural folds before migration. Adapted from (Couly et al., 1996).

B. Stylized right view of chicken embryo on embryonic day at E3.5, showing NC cell migration pathways in the craniofacial and branchial arch mesenchyme. Adapted from (Couly et al., 1998).

C. Distribution at E 8 of NC cells from r6-8 in cardiac outflow tract, including the tricuspid valves of aortic and pulmonary trunks, as well as pericytes and smooth muscle of blood vessels derived from PA3-6. Adapted from (Etchevers et al., 2001).

**Abbreviations**: AMes, anterior mesencephalon ; Ao, aorta; CC, common carotid arteries; Di, diencephalon; FN, frontonasal bud; LA, left atrium; LV, left ventricle; NC, neural crest; op, optic vesicle ; ot, otic vesicle; PMes, posterior mesencephalon ; PA1-6, pharyngeal arches 1-6 (by convention, there is no PA5 in amniotes); PT, pulmonary trunk; r1-8, rhombomeres 1-8; RA, right atrium; RV, right ventricle; SV, sinus venosus.

Etchevers, Dupin & Le Douarin

Table 1: The derivatives of the NC

| | Cell types | | | |
|---|---|---|---|---|
| | **Neurons and Glial cells** | **Pigment cells** | **Endocrine cells** | **Mesenchymal cells** |
| **CEPHALIC NC** | Sensory cranial ganglionic neurons | Skin melanocytes | Carotid body glomus cells | Cranio-facial skeleton |
| | Parasympathetic (eg., ciliary) ganglionic neurons | Inner ear and choroid melanocytes | Calcitonin-producing cells of the ultimobranchial body | *Dermal bone osteocytes* |
| | Enteric ganglionic neurons | Other extracutaneous melanocytes (in gums, meninges, heart…) | | *Endochondral osteocytes* |
| | Schwann cells along PNS nerves | | | *Chondrocytes* |
| | Satellite glial cells in PNS ganglia | | | |
| | Enteric glia | | | Other mesectodermal head and neck derivatives |
| | Olfactory ensheathing cells | | | *Myofibroblasts, smooth muscle cells and pericytes (cardiac outflow tract and pharyngeal arch-derived blood vessels)* |
| | | | | *Meninges (forebrain)* |
| | | | | *Cornea endothelial and stromal cells* |
| | | | | *Ciliary muscles/ anterior segment of the eye* |
| | | | | *Adipocytes* |
| | | | | *Facial dermis* |
| | | | | *Connective cells of glands, muscles and tendons* |
| | | | | *Odontoblasts, cells of periodontal ligament and tooth papillae* |
| **TRUNK NC** | Sensory (dorsal root) ganglionic neurons | Skin melanocytes | Adrenal medullary cells | |
| | Sympathetic ganglionic neurons | | | |
| | Parasympathetic ganglionic neurons | | | |
| | Satellite glial cells in PNS ganglia | | | |
| | Schwann cells along PNS nerves | | | |

| Year | 1868 | 1897 | 1950 | 1969-1978 | 1974 |
|---|---|---|---|---|---|
| Event | NC identified as a distinct structure in the chicken embryo | Evidence for the existence and functions of mesectoderm in the salamander | Experimental exploration of multiple amphibian NC lineages | Development of the quail-chick marking system to establish an exhaustive avian NC fate map | Introduction of the concept of neurocristopathies |
| References | His, 1868 | Platt, 1897 | Hörstadius, 1950 | Le Douarin, 1969, 1973; Le Douarin and Le Lièvre, 1970; Teillet and Le Douarin, 1970; Le Douarin et al., 1972; Le Douarin and Teillet, 1973, 1974; Le Lièvre and Le Douarin, 1973, 1975; Noden, 1978 | Bolande, 1974 |

| 1988 | 1988-2005 | 1993 | 1991-1996 | 2000 | 2009 |
|---|---|---|---|---|---|
| Single avian NC cells first shown to be multipotent *in vivo* | Evidence for avian and mammalian NC stem cell properties *in vitro* | Mapping of the three cell lineages, including NC, that contribute to the avian skull | Mutations in functionally conserved paracrine signaling pathways shown to underlie heritable human neurocristopathies | First lineage tracing of NC in mouse using genetic fate mapping tools (*Wnt1-Cre*) | Peripheral nerves shown to be embryonic and postnatal niches for Schwann cell-competent, multipotent NC progenitors |
| Bronner-Fraser and Fraser, 1988 | Baroffio et al., 1988, 1991; Stemple and Anderson, 1992; Morrison et al., 1999; Trentin et al., 2004; Kleber et al., 2005 | Couly et al., 1993 | Giebel and Spritz, 1991; Puffenberger et al., 1994; Hosoda et al., 1994; Edery et al., 1996; Hofstra et al., 1996 | Jiang et al., 2000, Chai et al., 2000 | Adameyko et al., 2009 |

2015

*In vivo* multipotency of mouse trunk NC cells (*R26-Confetti* lineage tracing)

Baggiolini et al., 2015

Etchevers et al, **Figure 1**

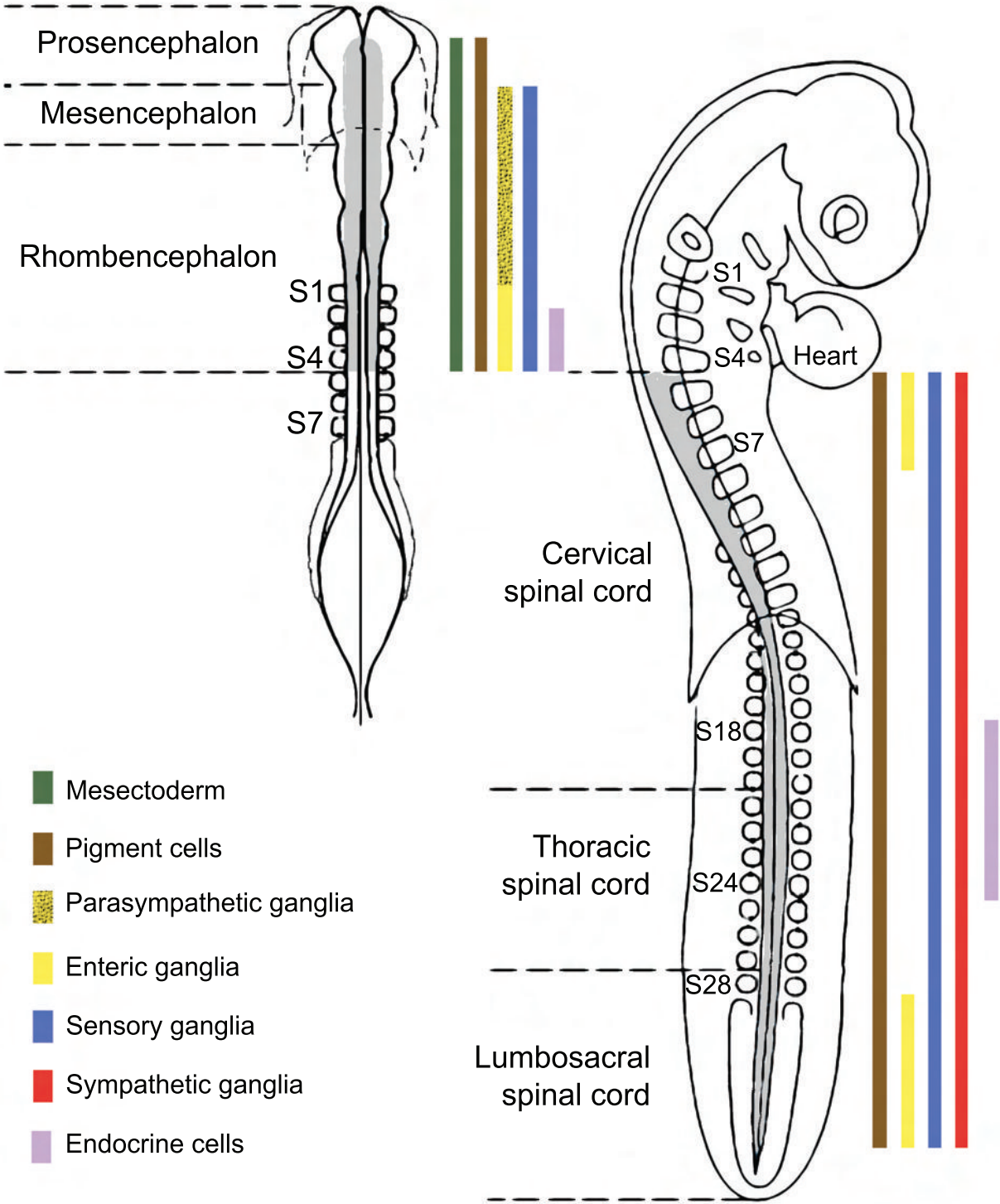

Etchevers et al, **Figure 2**

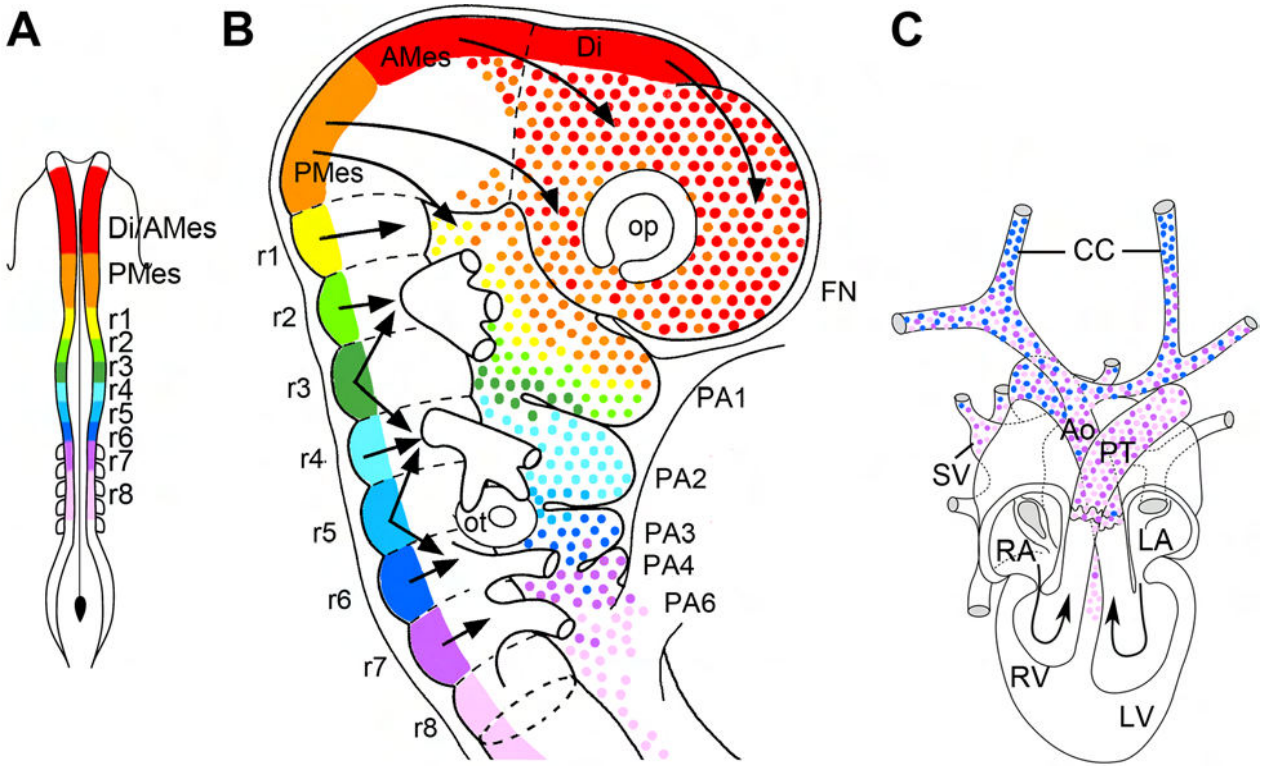